\documentstyle[aps,multicol,epsf]{revtex}

\draft

\begin{document}

\title{Self-Organized Branching Processes: A Mean-Field Theory for Avalanches}

\author{Stefano Zapperi, Kent B{\ae}kgaard Lauritsen,
	and H. Eugene Stanley}

\address{Center for Polymer Studies and Department of Physics,
	Boston University, Boston, Massachusetts 02215}

\date{\today}

\maketitle

\begin{abstract}
We discuss mean-field theories for self-organized criticality
and the connection with the general theory of branching processes.
We point out that the
nature of the self-organization is not addressed properly by the
previously proposed mean-field theories.
We introduce a new mean-field model that
explicitly takes the boundary conditions into account; in this way, the local
dynamical rules are coupled to a global equation that drives the
control parameter to its critical value.  We study the model numerically,
and analytically we compute the avalanche distributions.
\end{abstract}


%
%
%
%

\begin{multicols}{2}
\narrowtext

Many phenomena of interest to physicists, chemists, geologists, and biologists
display long-range spatiotemporal correlations.
The critical behavior is sometimes apparently ``spontaneous'',
in contrast to typical second order phase transitions, where scaling
occurs only when a parameter is tuned close to a critical value.
The concept of  self-organized criticality (SOC) \cite{btw} was introduced to
describe such systems which spontaneously organize into a critical state.

The SOC behavior has been investigated  using different models.
Among those the most widely studied, numerically
\cite{btw,nagel,grass} and analytically \cite{dhar,zap},
are the sandpile models. An important feature of such systems is that
they respond to external
perturbations by avalanches of all sizes $s$,
with a power-law distribution, $D(s)\sim s^{-\tau}$, where $\tau$
is a scaling exponent.

The simplest theoretical approach to SOC is  mean-field theory \cite{mf},
which allows for a qualitative description of the behavior of the system.
Mean-field exponents for SOC models have been obtained
by various approaches \cite{mf,dhar2,jl,fbs},
but it turns out that their values (e.g., $\tau=3/2$)
are the same for all the models considered thus far.
This fact can easily be understood since
the spreading of an avalanche in mean-field theory can be described by
a front consisting of non-interacting particles that can
either trigger subsequent activity or die out.
This kind of process is well known, and is called
a branching process \cite{harris}. The connection
between branching processes and SOC has been investigated,
and it has been proposed that the mean-field behavior of sandpile models
can be described by a {\em critical\/} branching process
\cite{alstr,crol,gp}.

We begin by noting the following paradox inherent in the previous mean-field
studies: for a branching process to be critical one must fine tune a
control parameter to a critical value. This, by definition, cannot
be the case for a SOC system, where the critical state is approached
dynamically without the need to fine tune any parameter.
In the present paper, we resolve this paradox by introducing
a new mean-field model, the self-organized branching-process (SOBP).
The coupling of the local dynamical rules to a global
condition, drives the system into a state that
is indeed described by a critical branching process. We study the model
numerically and compare the results of the simulations
with analytical results.
We introduce the SOBP model based on physical considerations,
and then show that the mean-field theory of SOC models can be exactly mapped
to the SOBP model.

Sandpile models are cellular automata defined on a $d-$dimensional
lattice. An integer variable, which we call energy, is associated with
each site of the lattice. Over time, energy is added in integer units
to a randomly chosen site in the system.
Whenever the energy on a site reaches a critical value, the site ``relaxes''
and its energy is transferred to the neighbors according to the
specific rules of the model. In this way, a single relaxation can trigger other
relaxations, leading to the formation of an avalanche.
With {\em closed\/} boundary conditions, the density of critical sites
in the system will increase as a function of time
such that eventually an avalanche of infinite lifetime will form.
Therefore, in order for the system to be able to reach a SOC state,
it is essential to impose
{\em open\/} boundary conditions such that energy can leave the system,
thus decreasing the number of critical sites.
As a result, regardless of the initial conditions, the system is
driven to a stationary SOC state characterized by a balance between input
and output, i.e., on average the energy added to the system is equal
to the energy that leaves the system through the boundaries.

In the mean-field description of the sandpile model ($d \to \infty$)
one neglects correlations, which implies that
avalanches do not form loops and hence spread as a
branching process.
In the SOBP model, an avalanche starts with a single active site, which then
relaxes with probability $p$, leading to two new active sites. With
probability $1-p$ the initial site does not relax and the avalanche stops.
If the avalanche does not stop,
we repeat the procedure for the new active sites until no active site remains.
The parameter $p$ is the probability that a site relaxes
when it is triggered by an external input.
For the SOBP branching process, there is a critical value, $p_c=1/2$,
such that for  $p>p_c$ the
probability to have an infinite avalanche is non-zero, while for
$p<p_c$ all avalanches are finite. Thus, $p=p_c$ corresponds
to the critical case, where avalanches are power law distributed.

\begin{figure}[htb]
\centerline{
        \epsfxsize=8.0cm
	\epsfbox{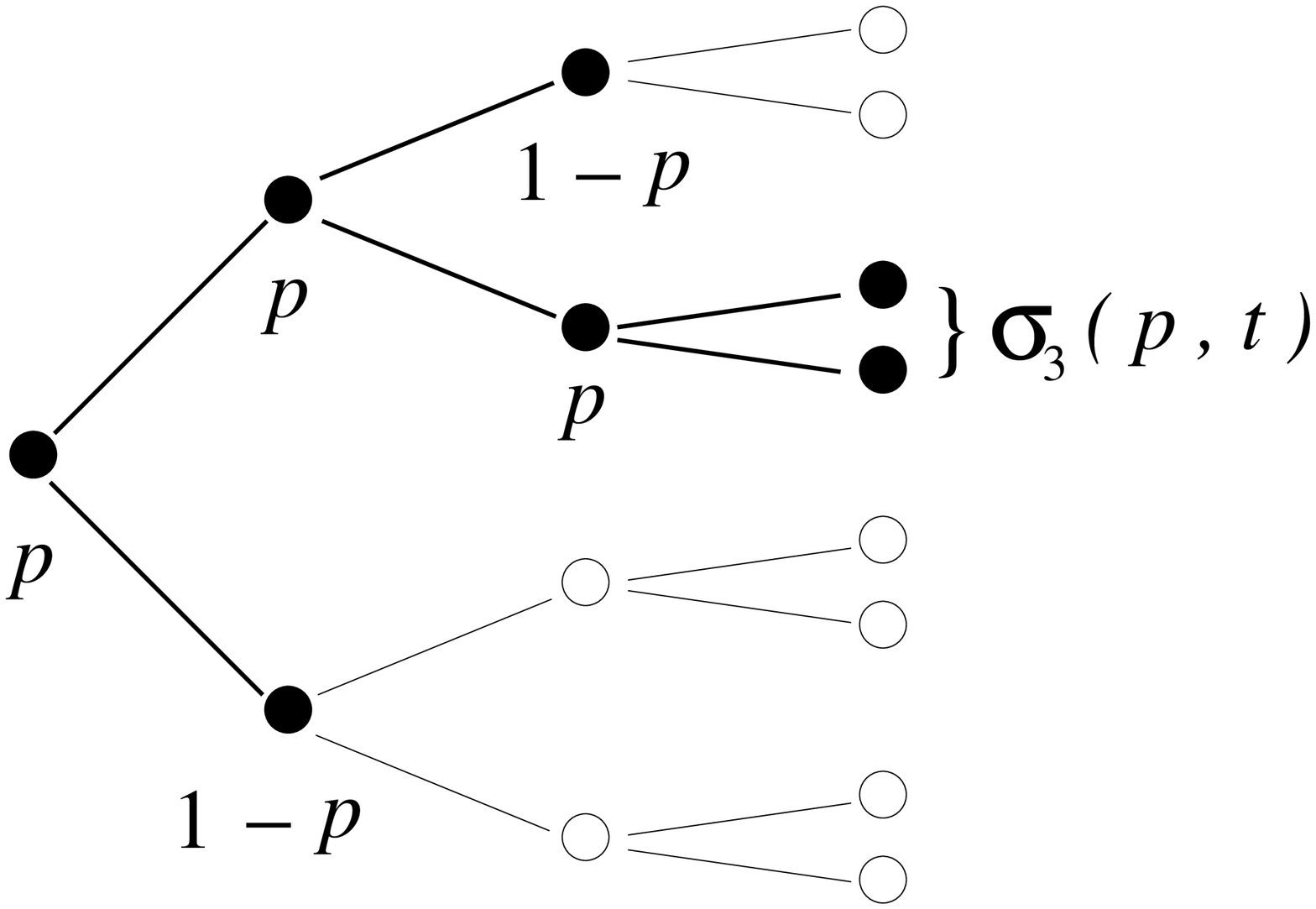}
        \vspace*{0.5cm}
	}
\caption{Schematic drawing of an avalanche in a system with a
	maximum of $n=3$ avalanche generations corresponding to
	$N=2^{n+1}-1=15$ sites. Each black site relaxes with probability $p$
	to two new black sites and with probability $1-p$ to two white sites.
	The black sites are part of an avalanche of size $s=7$,
	whereas the active sites at the boundary yield $\sigma_3(p,t)=2$.
	Note that $s$ in the SOBP model must be an odd integer
	while $\sigma_n$ will be even.
	}
\label{fig:1}
\end{figure}

In the above description, however,
the boundary conditions are not taken into account---even though
they are crucial for the self-organization process.
We can introduce boundary conditions in the problem in a natural way,
by allowing for no more than $n$ generations for each avalanche.
Schematically, we can view the evolution of a single
avalanche of size $s$ as taking place on a tree
of size $N=2^{n+1}-1$ (see Fig.~\ref{fig:1}).
If the avalanche reaches the boundary of the tree,
we count the number of active sites $\sigma_n$
(which in the sandpile language corresponds to the energy leaving the system),
and we expect that $p$
decreases for the next avalanche. If, on the other hand, the avalanche
stops before reaching the boundary, then $p$ will slightly increase.
Note that we are not studying the model on a
Bethe lattice, i.e., the branching structure we are discussing
is not directly related to the geometry of the system.
The number of generations $n$ can, nevertheless, be thought of as some
measure of the linear dimension of the system.

The scenario discussed above is described by the following
dynamical equation for $p(t)$:
\begin{equation}
	p(t+1)=p(t)+\frac{1-\sigma_n(p,t)}{N} .
                                                \label{eq:p(t)-disc}
\end{equation}
Here $\sigma_n$, the size of an avalanche reaching the boundary,
fluctuates in time and hence acts as a stochastic driving force.
If $\sigma_n=0$, then $p$ increases
(because some energy has been put into the system without any output),
whereas if $\sigma_n>0$ then $p$ decreases (due to energy leaving the system).
Equation~(\ref{eq:p(t)-disc})
describes the global dynamics of the SOBP, as opposed
to the local dynamics which is given by the branching process.
We can study the model for a fixed value of $n$, and then take
the limit $n\to \infty$. In this way, we perform the long-time limit before
the ``thermodynamic'' limit,
which corresponds exactly to what is done in sandpile simulations.

The SOBP model can be exactly mapped to SOC models
in the limit $d \to \infty$, i.e., it provides a mean-field theory
of self-organized critical systems.
To show this, we consider for simplicity the two-state sandpile model
\cite{manna}: When a particle is added to a site $z_i$,
the site will relax if $z_i=1$.
In the limit $d \to \infty$, the avalanche will never visit the same
site more than once. Accordingly, each site in the avalanche
will relax with the same probability $p=P(z=1)$.
Eventually, the avalanche will stop, and $\sigma \ge 0$ particles will
leave the system.
Thus, the total number of particles $M(t)$ evolves according to
\begin{equation}
	M(t+1) = M(t) + 1 - \sigma .
						\label{eq:M(t)}
\end{equation}
The dynamical equation~(\ref{eq:p(t)-disc})
for the SOBP model is recovered by noting that $M(t)=N P(z=1) = Np$.
For other SOC systems,
mean-field descriptions similar to the SOBP model can be derived.

Before investigating the SOBP model analytically, we
present some preliminary considerations together with numerical results.
For a fixed value of $p$, the average value of $\sigma_n$ is 
$
	\left< \sigma_n(p,t) \right> =(2p)^n \cite{harris} .
$
We can write
$
	\sigma_n(p,t)=(2p)^n + \eta(p,t),
$
where the ``noise'' $\eta$ describes the fluctuations around the average.
Inserting  this expression in Eq.~(\ref{eq:p(t)-disc}) and taking the
continuum time limit, we obtain
\begin{equation}
	\frac{d p}{d t}=\frac{1-(2p)^n}{N}+\frac{\eta(p,t)}{N}.
                                                \label{eq:p(t)-cont}
\end{equation}
Without the last term, Eq.~(\ref{eq:p(t)-cont})
has a fixed point (${d p}/{d t}=0$) for $p=p_c=1/2$. On linearizing
Eq.~(\ref{eq:p(t)-cont}), we see that the fixed point is
attractive, which demonstrates the self organization of the SOBP model since
the noise $\eta/N$ will have vanishingly small effect in the
thermodynamic limit.  To confirm our statement,
we have studied the SOBP model by carrying out simulations for different
system sizes, and averaging over typically $5\times 10^6$ realizations.

\begin{figure}[htb]
\centerline{
        \epsfxsize=8.0cm
	\epsfbox{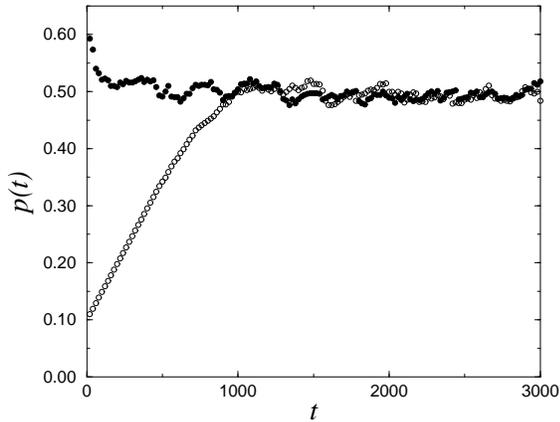}
	}
\caption{The value of $p$ as a function of time for a system with $n=10$
	generations. The two curves refer to two different initial conditions,
	above $p_c$ ($\bullet$) and below $p_c$ ($\circ$).
	After a transient, the control parameter $p(t)$ reaches its
	critical value $p_c$ and fluctuates around it with
	short-range correlations.
	}
\label{fig:2}
\end{figure}

In Fig.~\ref{fig:2} we show the value of $p$ as a function of time.
Independent of the initial conditions, we find
that after a transient $p(t)$ reaches the self-organized state
described by the critical value
$p_c=1/2$ and fluctuates around it with short-range correlations
(of the order of one time unit).
By computing the variance of $p(t)$,
we find that the fluctuations around the critical value
decrease with the system size as $1/N$.
Moreover, by analyzing the higher moments, we find that the fluctuations
for $N \gg 1$ can be very well described by a Gaussian distribution,
\begin{equation}
        \phi(p) \equiv \frac{1}{\sqrt{2\pi\Delta_N}\,}
		\exp\left( -\frac{(p-p_N)^2}{2\Delta_N} \right)   .
                                                \label{eq:phi(p)}
\end{equation}
Here $p_N \equiv p_c - {a}/{N}$, $\Delta_N \equiv {b}/{N}$,
with $a = 0.69 \pm 0.02$, and $b = 0.26 \pm 0.01$.
In the limit $N \to \infty$, the distribution $\phi(p)$ therefore approaches
a delta function, $\delta(p-p_c)$.

\begin{figure}[htb]
\centerline{
        \epsfxsize=8.0cm
	\epsfbox{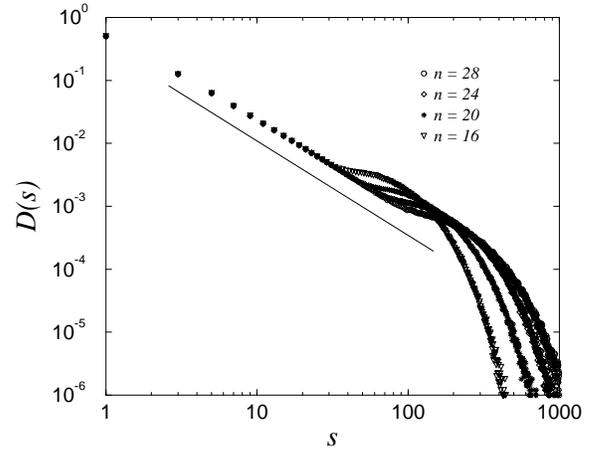}
	}
\caption{Log-log plot of the avalanche distribution $D(s)$
	for different system sizes.
	The number of generations $n$ increases from left to right.
	A line with slope $\tau=3/2$ is plotted for reference,
	and it describes the behavior of the data for intermediate $s$ values,
	cf.\ Eq.~(\protect\ref{eq:n(s)-limit}).
	For large $s$, the distributions fall off exponentially.
	}
\label{fig:3}
\end{figure}

\begin{figure}[htb]
\centerline{
        \epsfxsize=8.0cm
	\epsfbox{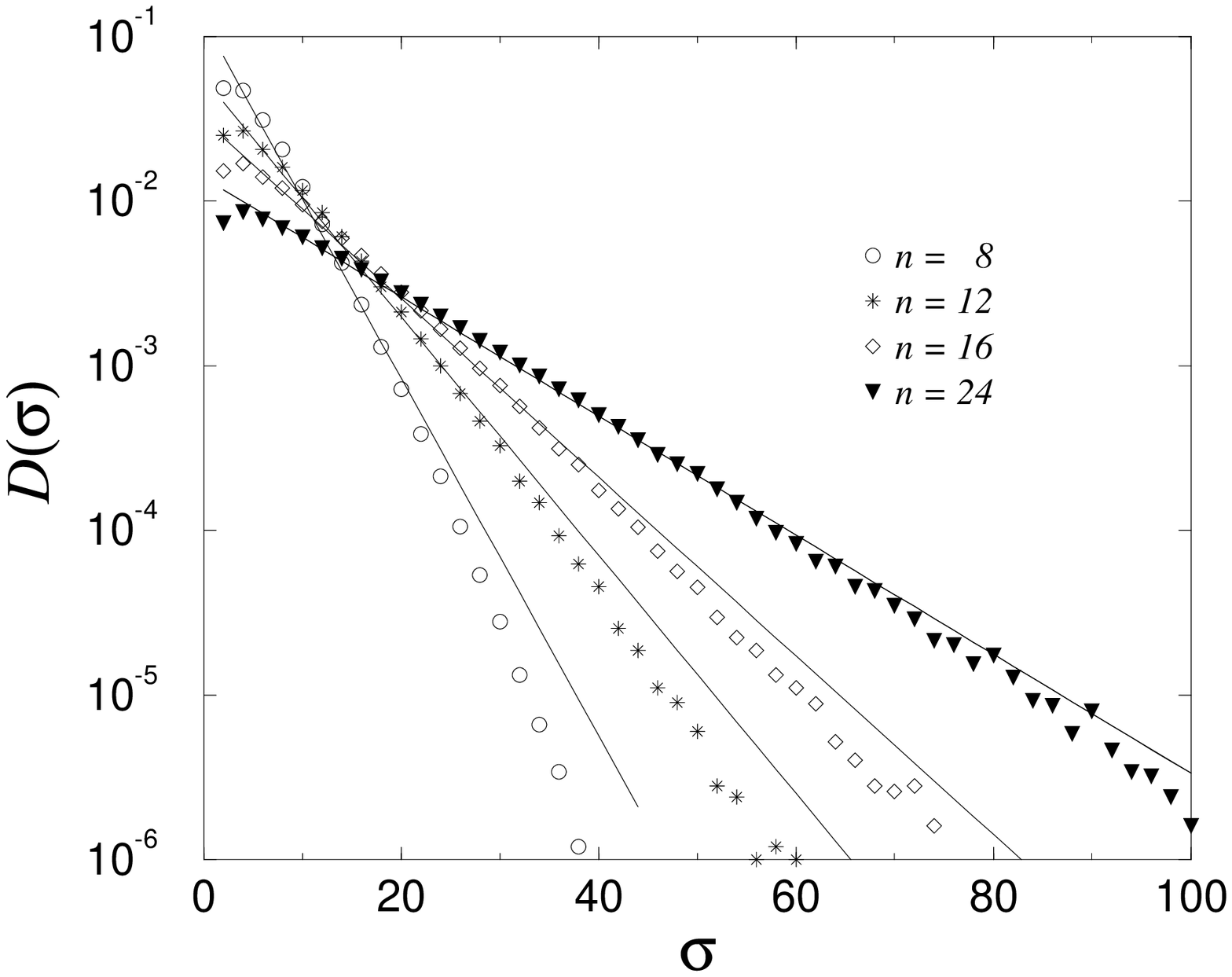}
	}
\caption{Semilogarithmic plot of the boundary avalanche distribution
	$D(\sigma)$ for different system sizes \protect\cite{sigma=0-peak}.
	For large $\sigma$, the distributions fall off exponentially.
	The solid lines are our analytical results of
	Eq.~(\protect\ref{eq:n(sigma)}) for $n \gg 1$.
	We note that the agreement improves with increasing $n$.
	}
\label{fig:4}
\end{figure}

In Fig.~\ref{fig:3}
we show the avalanche size distribution $D(s)$ for different values of
the number of generations $n$. We observe a scaling region
($D(s)\sim s^{-\tau}$ with $\tau=3/2$),
whose size increases with $n$, and an exponential cutoff. The power-law
scaling is a signature of the mean-field criticality of the SOBP model.
In Fig.~\ref{fig:4} we report the distribution of active sites at the
boundary, $D(\sigma)$, for different values of the number of
generations. This distribution falls off exponentially.

We now present some analytical results in the limit where $n \gg 1$.
First, we discuss the branching process for any value of $p$, and show
how this for the critical branching process with $p=p_c$ yields
the mean-field exponent $\tau=3/2$. In addition, we can obtain
results for finite, but large, values of $n$.
Next, we obtain results for the SOBP model which, as shown in
Fig.~\ref{fig:2}, self-organizes into a critical state where $p=p_c$.

We introduce the probabilities
$P_n(s,p)$ and $Q_n(\sigma,p)$ of having an avalanche of size
$s$ and boundary size $\sigma$ in a system with $n$ generations.
The corresponding generating functions are defined by \cite{harris}
\begin{mathletters}
\label{eq:hg_def}
\begin{equation}
        f_n(x,p) \equiv \sum_{s} P_n(s,p) x^s   ,
						\label{eq:f_n-def}
\end{equation}
\begin{equation}
	g_n(x,p) \equiv \sum_{\sigma} Q_n(\sigma,p) x^\sigma   .
						\label{eq:g_n-def}
\end{equation}
\end{mathletters}
Due to the hierarchical structure of the branching process, it is
possible to write down recursion relations for $P_n(s,p)$ and
$Q_n(\sigma,p)$, from which we obtain \cite{harris}
\begin{mathletters}
\label{eq:fg_n+1}
\begin{equation}
        f_{n+1}(x,p) = x \left[ (1-p) + p f^{2}_{n}(x,p) \right]  ,
                                                \label{eq:f_n+1}
\end{equation}
\begin{equation}
        g_{n+1}(x,p) = (1-p) + p g^{2}_{n}(x,p)  ,
                                                \label{eq:g_n+1}
\end{equation}
\end{mathletters}
where $f_0(x,p)=g_0(x,p)=x$.

Next, we calculate the avalanche distribution $D(s)$ determined by
$P_n(s,p)$ by using the recursion relation~(\ref{eq:f_n+1}).
The solution of Eq.~(\ref{eq:f_n+1}) in the limit $n \gg 1$ is given by
\begin{equation}
	f(x,p) = \frac{1-\sqrt{1-4x^2p(1-p)} \, }{2xp} .
                                                \label{eq:f*}
\end{equation}

By expanding Eq.~(\ref{eq:f*}) as a series in $x$, and comparing with
the definition~(\ref{eq:f_n-def}), we obtain for sizes such that
$1 \ll s \lesssim n$ \cite{small-s}
\begin{equation}
        P_n(s,p) = \frac{\sqrt{2(1-p)/\pi p}}{s^{3/2}} \,
		   \exp\left( -s / s_c(p) \right)  .
                                                         \label{eq:P_n}
\end{equation}
The cutoff $s_c(p)$ is given by \mbox{$s_c(p) = -2/\ln 4p(1-p)$}.
As \mbox{$p\to 1/2$}, \mbox{$s_c(p)\to \infty$},
thus showing explicitly that the critical value for the
branching process is $p_c=1/2$, and that the mean-field exponent
for the critical branching process is $\tau=3/2$.

The expression~(\ref{eq:P_n}) is only valid for avalanches
which do not feel the finite size of the system.
For avalanches with $n \lesssim s \lesssim N$,
it is possible to solve the recursion
relation~(\ref{eq:f_n+1}), and then obtain $P_n(s,p)$ for $p \ge p_c$
by the use of a Tauberian theorem
\cite{feller,asmussen-hering,weiss}.
By carrying out such an analysis, we obtain after some algebra
\begin{equation}
        P_n(s,p) \approx
		 A(p) \, \exp\left( -s/s_0(p) \right) ,
                                                \label{eq:P_n(s)-tail}
\end{equation}
with functions $A(p)$ and $s_0(p)$ which can not be
determined analytically. Nevertheless, we see that for any $p$
the probabilities $P_n(s,p)$ will decay exponentially.

The final step is to calculate the avalanche distribution
$D(s)$ for the SOBP model. This can be calculated
as the average value of $P_n(s,p)$ with respect to the probability
density $\phi(p)$, i.e., according to the formula
\begin{equation}
        D(s) = \int_{0}^{1} dp \, \phi(p) \, P_n(s,p)   .
                                                \label{eq:n(s)}
\end{equation}
The simulation results in
Fig.~\ref{fig:2} show that $\phi(p)$ for $N \gg 1$ approaches
the delta function $\delta(p-p_c)$ [cf.\ Eq.~(\ref{eq:phi(p)})].
Thus, for $1 \ll s \lesssim n$, we obtain the power-law behavior
\begin{equation}
        D(s) = \sqrt{\frac{2}{\pi}\,} \, s^{-\tau}  ,
                                                \label{eq:n(s)-limit}
\end{equation}
where $\tau=3/2$,
and for $s \gtrsim n$ we obtain an exponential cutoff $\exp(-s/s_0(p_c))$.
These results are in good agreement with our numerical results shown in
Fig.~\ref{fig:3}.
The deviations from the power-law behavior~(\ref{eq:n(s)-limit})
are due to the fact that Eq.~(\ref{eq:P_n}) is only valid for
$1 \ll s \lesssim n$ \cite{small-s}.
Performing a more accurate calculation,
by the use of the Gaussian distribution~(\ref{eq:phi(p)})
instead of the $N \to \infty$ limit, $\delta(p-p_c)$,
we recover the power-law behavior in Eq.~(\ref{eq:n(s)-limit})
with correction terms.

In an analogous fashion, we can calculate the asymptotic form of
$Q_n(\sigma,p)$ for $1 \ll \sigma \lesssim n$ and $p \ge p_c$ by the use of
a Tauberian theorem, with the result
\begin{equation}
        Q_n(\sigma,p) \approx \frac{2(2p)^n}{\sigma_0(1+\sigma_0)} \,
				 \exp\left( -\sigma/\sigma_0(p) \right) ,
                                                \label{eq:P_n(si)-tail}
\end{equation}
where
$
	\sigma_0(p) \equiv \left( (2p)^n - 1 \right) / 2\ln 2p
$.
Hence, in the SOBP model, the boundary avalanche distribution is
\begin{equation}
        D(\sigma) = \int_{0}^{1} dp \, \phi(p) \, Q_n(\sigma,p)
		   = \frac{8}{n^2} \, \exp\left( -2\sigma/n \right) ,
                                                \label{eq:n(sigma)}
\end{equation}
which agrees with our simulation results for $n \gg 1$
as shown in Fig.~\ref{fig:4} \cite{sigma=0-peak}.

The avalanche lifetime distribution
$
	L(t) \sim t^{-y}
$
can also be computed. {}From $D(\sigma_m=0)=1-2/m$ \cite{sigma=0-peak}
for a system with $m$ generations, we obtain
$L(m) \sim m^{-2}$. Identifying the number of generations $m$ of an
avalanche with the time $t$, we thus obtain the mean-field value $y=2$.
This result is in complete agreement with simulations on the SOBP model.

In summary, we have introduced a self-organized
branching process (SOBP) that captures the physical features
of the self-organization mechanism in sandpile models.
Previous investigations have focused on the critical branching process
as the mean-field description of sandpile systems.
In the SOBP model, however,
by explicitly incorporating the boundary conditions,
we were able to show, using Eq.~(\ref{eq:p(t)-cont}),
how the dynamics
drives the system into a stationary state, which in the thermodynamic
limit corresponds to the critical branching process.
We have calculated the avalanche distributions
analytically, and found the mean-field result $\tau=3/2$,
in good agreement with our simulation results.

We thank S.~V.~Buldyrev and R.~Sadr for discussions concerning the
calculations, and L.~Amaral, R.~Cuerno, B.~Kutnjak-Urbanc, H.~Makse,
and S.~Milo\v sevi\'c for comments on the manuscript.
K.~B.~L.\ acknowledges the support from the Danish Natural Science
Research Council. The Center for Polymer Studies is supported by NSF.

\end{multicols}

\twocolumn

\vspace*{-0.5cm}


\begin{thebibliography}{99}

\vspace*{-1.5cm}




\bibitem{btw}
	P. Bak, C. Tang, and K. Wiesenfeld,
	Phys. Rev. Lett. {\bf 59}, 381 (1987);
 	Phys. Rev. A {\bf 38}, 364 (1988).
\bibitem{nagel}
 	L. P. Kadanoff, S. R. Nagel, L. Wu, and S. Zhu,
	Phys. Rev. A {\bf 39}, 6524 (1989).
\bibitem{grass}
 	P. Grassberger and S. S. Manna,
	J. Phys. France {\bf 51}, 1077 (1990).
\bibitem{dhar}
 	D. Dhar and R. Ramaswamy, Phys. Rev. Lett. {\bf 63}, 1659 (1989);
 	D. Dhar, {\it ibid.\/} {\bf 64}, 1613 (1991);
 	S. N. Majumdar and D. Dhar, Physica A {\bf 185}, 129 (1992).
\bibitem{zap}
 	L. Pietronero, A. Vespignani, and S. Zapperi,
	Phys. Rev. Lett. {\bf 72}, 1690 (1994);
	Phys. Rev. E {\bf 51}, 1711 (1995).
\bibitem{mf}
	C. Tang and P. Bak, J. Stat. Phys. {\bf 51}, 797 (1988).
\bibitem{dhar2}
	D. Dhar and S. N. Majumdar, J. Phys. A {\bf 23}, 4333 (1990).
\bibitem{jl}
	S. A. Janowsky and C. A. Laberge,
	J. Phys. A {\bf 26}, L973 (1993).
\bibitem{fbs}
	H. Flyvbjerg, K. Sneppen, and P. Bak,
	{Phys. Rev. Lett.} {\bf 71}, 4087 (1993);
	J. de Boer, B. Derrida, H. Flyvbjerg, A. D. Jackson, and T. Wettig,
	{\it ibid.} {\bf 73}, 906 (1994).
\bibitem{harris}
        T. E. Harris,
        {\em The Theory of Branching Processes} (Dover, New York, 1989).
\bibitem{alstr}
	P. Alstr\o m,
	Phys. Rev. A {\bf 38}, 4905 (1988).
\bibitem{crol}
	K. Christensen and Z. Olami,
	Phys. Rev. E {\bf 48}, 3361 (1993).
\bibitem{gp}
        R. Garc\'\i a-Pelayo,
        {Phys. Rev.} E {\bf 49}, 4903 (1994).
\bibitem{manna}
	S. S. Manna, J. Phys. A {\bf 24}, L363 (1991):
	In this two-state model, the energy takes the two stable values,
	$z_i=0$ (empty) and $z_i=1$ (particle). When $z_i \ge z_c$,
	with $z_c=2$, the site relaxes by distributing two particles
	to two randomly chosen neighbors.
\bibitem{small-s}
	For small $s$ we have the exact results
	$P_n(1,p)=1-p$, $P_n(3,p)=p(1-p)^2$, $P_n(5,p)=2p^2(1-p)^3$,
	and so forth.
\bibitem{feller}
	W. Feller,
	{\em An Introduction to Probability Theory and its Applications},
	Vol.~2, 2nd ed.\ (John-Wiley, New York, 1971).
\bibitem{asmussen-hering}
        S. Asmussen and H. Hering,
        {\em Branching Processes} (Birkh\"auser, Boston, 1983).
\bibitem{weiss}
	G. H. Weiss,
	{\em Aspects and Applications of the Random Walk}
	(North-Holland, Amsterdam, 1994).
\bibitem{sigma=0-peak}
	$D(\sigma=0)=1-2/n$ (cf.\ \protect\cite{harris}).



\end{thebibliography}
\end{document}